\documentstyle[aps, epsfig, multicol, prl]{revtex}

\begin{document}
\draft
\title{Breakdown of correspondence in chaotic systems:\\
 Ehrenfest versus localization times}
\author{Zbyszek P. Karkuszewski$^{1,2}$, Jakub Zakrzewski$^2$
 and Wojciech H. Zurek$^1$ }

\address{
        $^1$Theoretical Division, T6, MS B288, LANL, Los Alamos, NM 87545\\
        and\\
        $^2$Instytut Fizyki, Uniwersytet Jagiello\'nski, Reymonta 4, 30-059 Krak\'ow, Poland}

\date{\today}

\maketitle
\begin{abstract}
Breakdown of quantum-classical correspondence is studied on an experimentally
realizable example of one-dimensional periodically driven system. 
Two relevant time scales are identified in this system: 
the short Ehrenfest time $t_{\hbar}\sim \ln(\hbar^{-1})$
and the typically much longer localization time scale $T_L\sim \hbar^{-2}$.
It is shown that surprisingly weak modification of the Hamiltonian may
eliminate the more dramatic symptoms of localization without effecting
the more subtle but ubiquitous and rapid loss of correspondence at $t_{\hbar}$.
\end{abstract}

\begin{multicols}{2}
Quantum realizations of classically
chaotic systems behave differently than their fully classical counterparts.
The breakdown of quantum-classical correspondence occurs on two distinct
time scales: the short Ehrenfest time
\begin{equation}
t_{\hbar}\cong \frac{1}{\lambda}\ln\left(\frac{\Delta p \chi}{\hbar}\right)
\cong \frac{1}{\lambda}\ln\left(\frac{A}{\hbar}\right)
\label{e1}
\end{equation}
gives the time scale at which the quantum minimal wavepacket spreads sufficiently
over a macroscopic part of the phase-space to feel nonlinearities in the
potential \cite{berman78,zaslavsky81,r2}. In (\ref{e1}) $\Delta p$ denotes
the initial uncertainty in momentum, $\lambda$ is the Lyapunov
exponent, $\chi\cong\sqrt{\partial_x V/\partial^3_x V}$ is a typical scale
on which nonlinearities in the potential $V$ are significant.
Ehrenfest time $t_{\hbar}$ marks the moment when quantum corrections
become important in the time evolution of the Wigner function $W(x,p)$,
governed by the Wigner transform of the von Neumann bracket, known as the
Moyal bracket \cite{Wigner}:
\begin{eqnarray}
\label{Moyal}
\frac{d}{dt}W &=& \{H,W\}_{MB}\\
{}&=& \{H,W\} +
\sum\limits_{n=1}^\infty\frac{(-1)^n\hbar^{2n}}{2^{2n}(2n+1)!}
\frac{\partial^{2n+1}V}{\partial x^{2n+1}}
\frac{\partial^{2n+1}W}{\partial p^{2n+1}}, \nonumber
\end{eqnarray}
The Moyal bracket appearing above is defined as
$$
\{H,W\}_{MB}=-\frac{2}{\hbar}
H\sin\left( \frac{\hbar}{2}\left(\frac{\buildrel\leftarrow\over\partial}
{\partial p}\frac{\buildrel\rightarrow\over\partial}{\partial x}
-\frac{\buildrel\leftarrow\over\partial}{\partial x}
\frac{\buildrel\rightarrow\over\partial}{\partial p}\right)\right)W.
$$
The arrows indicate the direction of differentiation.

In a chaotic system initially localized distribution will be stretched in the
unstable directions. In a Hamiltonian system, this stretching must
be matched by shrinking in the stable directions as mandated by the Liouville
theorem. Consequently, the typical width of the effective support of
classical probability distribution shrinks as $\exp(-\lambda t)$.
That leads to an exponential increase of higher order derivatives of $W$
in (\ref{Moyal}), so after $t_{\hbar}$ quantum corrections become comparable
to terms given by the Poisson bracket \cite{r2}.

There is evidence that at $t_{\hbar}$ also the averages over the phase
space show symptoms of correspondence loss \cite{r3,r4} although not
everyone agrees \cite{r5,r6}. The first goal of our paper is then
to investigate the dependence of the onset of the loss of correspondence
in the averages on $\hbar$, and to show that it is consistent with (\ref{e1}).

In addition to the rapid correspondence loss on the logarithmic time scale
$t_{\hbar}$ there is plentiful evidence for the dynamical localization in
a class of quantum chaotic systems \cite{r7,r8,Raizen}. It sets in on a much
longer localization time scale:
\begin{equation}
T_L\cong \frac{d}{\hbar^2}
\label{e3}
\end{equation}
where $d$ is diffusion constant characterizing initial growth of variance in
momentum of a quantum system. Tutorial example of localization
is described in \cite{Haake}.
It is of obvious interest to consider systems in which both mechanisms of
correspondence loss are present, and to investigate their behavior with
varying parameters. The system we investigate is a straightforward modification
of experimentally studied system based on an optical lattice. Therefore, 
it should be
 simple to verify our predictions in the laboratory.

We study quantum-classical correspondence in a system described by the
Hamiltonian
\begin{equation}
H(x,p) = \frac{p^2}{2m} -\kappa\cos(x-l\sin t) +a\frac{x^2}{2}.
\label{e5}
\end{equation}
For $a=0$ this Hamiltonian reduces to a model recently investigated
both theoretically \cite{Shepelyansky,Zoller,Bialynicki} and
experimentally \cite{Raizen} corresponding to motion of cold atoms
in a quasi-resonant standing wave with periodically moving nodal pattern.
Classical behavior of the system is of the mixed type with large regions
of phase space dominated by a chaotic motion (e.g. Lyapunov exponent of
the chaotic component is $\lambda=0.2$ for chosen parameter values
$m=1,\ \kappa=0.36,\ l=3,\ a=0$, \cite{Zoller}).

To investigate quantum-classical correspondence one can start by
looking at the Wigner function. Structures which appear in $W$ are
known to be different from (but not unrelated to) the classical probability
distribution in the phase space. One can especially note existence
of the interference fringes that saturate on small scales associated
with action $s\sim (2\pi\hbar)^2/S$, where $S$ is the action of the system
-- e.g., the phase space volume over which $W$ can spread \cite{r10}.

One might nevertheless suppose that the quantum structures appearing on such 
small in the Wigner function will have little effect on the averages. 
In particular, averages typically
concern large scales, and, therefore, should not be directly
effected by small-scale interference. This is, however, only partially true.

\begin{figure}[htb]
\centering
{\epsfig{file=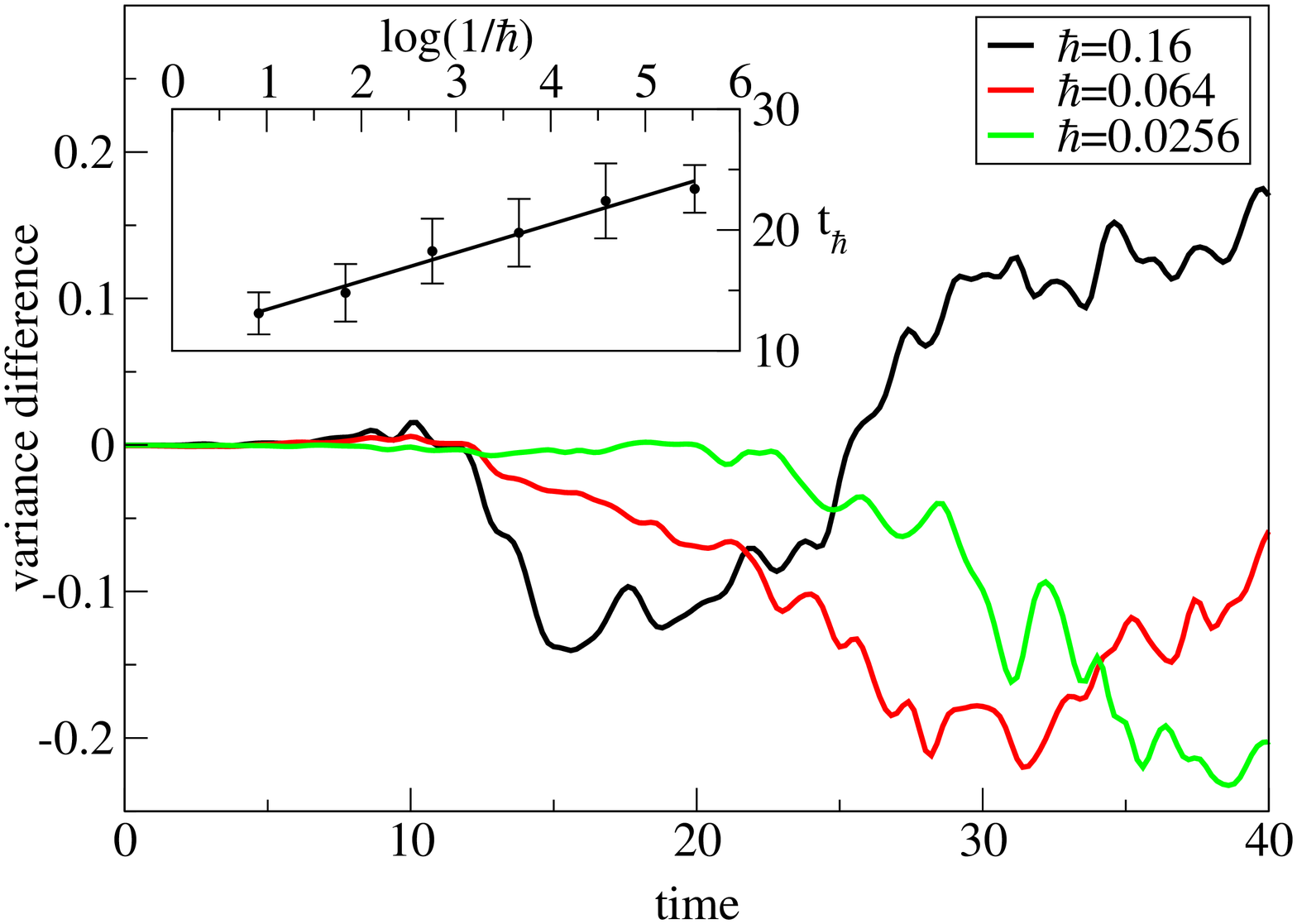, width=8.6cm}}
\caption{Difference between classical, $\Delta^2_{\mathrm cl}$,
 and quantum, $\Delta^2_{\mathrm qm}$, variances for three values
of $\hbar$ and for the same initial condition i.e. coherent state centered at
 (0.0) in the phase space. In the insert, $t_{\hbar}$ dependence on
$\log(1/\hbar)$ averaged over few initial conditions is shown. Error bars
indicate standard deviation of the results for given value of $\hbar$.
Criterion used to estimate Ehrenfest time scale compares magnitude of
Poisson bracket and a quantum correction (the rest of the Moyal bracket in
(\ref{Moyal})).} 
\label{fig1}
\end{figure}

\begin{figure}[htb]
\centering
{\epsfig{file=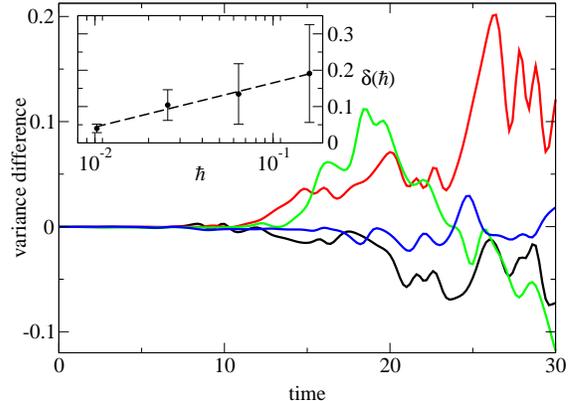, width=8.6cm}}
\caption{Difference between classical and quantum variances for $\hbar= 0.0256$
and four initial conditions [coherent states  centered at
points (0.0), ($\pi/2$, 0), ($-\pi/2$, 0) and ($\pi$, 0) in the phase space]
is plotted
using black, red, green and blue lines respectively. The insert shows the
discrepancy between classical and quantum averages as defined by
(\protect{\ref{discrepancy}}) as a function of $\hbar$. The dependence seems
to be logarithmic.}
\label{fig2}
\end{figure}

Figure \ref{fig1} shows the difference between  classical, 
$\Delta^2_{\mathrm cl}$,
 and quantum, $\Delta^2_{\mathrm qm}$,
averages of a typical quantity ($\Delta^2 =\langle p^2\rangle -
\langle p\rangle^2$) for $\hbar = 0.16$, 0.064 and 0.0256 obtained
in the time evolution resulting from Hamiltonian (\ref{e5}) with $a=0$.
Initial classical distribution is a two-dimensional Gaussian in phase
space and corresponds to a Gaussian wavepacket used for quantum
evolution.
The loss of correspondence is apparent: early on, a noticeable difference between
the two averages develops. The smaller the value of $\hbar$ ,
the later it appears.
The insert shows that the discrepancies occur after a time which seems
indeed logarithmic in $\hbar$. They persist without
 a significant change in
their magnitude until localization (which occurs
at a $T_L\sim \hbar^{-2}$, a time scale much more sensitive to changes of
$\hbar$ than the logarithmic $t_{\hbar}$) begins to dominate.
Vertical error bars in the insert arise from the fact that
the local Lyapunov exponents (i.e. evaluated over a time scale of order $t_\hbar$, (\ref{e1})), differ for various initial
conditions at given $\hbar$. This in turn affects value of $t_\hbar$ (see
Fig.~\ref{fig2}).

It is intriguing to enquire about the magnitude of the typical discrepancy
between the quantum and classical averages in the time interval between
$t_\hbar$ and $T_L$. To address this question we have evaluated
\begin{equation}
        \delta(\hbar)= \frac{1}{t_2-t_1}\int_{t_1}^{t_2}
        |\Delta^2_{\mathrm  cl}-\Delta^2_{\mathrm qm}| \mbox{d}t,
\label{discrepancy}
\end{equation}
for $t_\hbar<t_1 <t_2 <T_L$ as a function of $\hbar$.
 The results (averaged over several initial conditions)
show that $\delta(\hbar)$ decreases with decreasing $\hbar$ significantly
slower than linearly (compare the insert in Fig~\ref{fig2}).
 While our data do not allow for a definite conclusion, they 
point out towards a logarithmic $\hbar$ dependence of $\delta(\hbar)$.
The results obtained for $\delta(\hbar)$ depend slightly on the
choice of $t_1$ and $t_2$. That, together with the size of the error bars,
does not allow us to be more definitive.
The logarithmic dependence is not in disagreement with results of \cite{EB} 
(who claim that largest deviations are ${\cal O}(\hbar)$) obtained for a model 
of two interacting spins. Our data indicate a similar behavior for a realistic,
experimentally accessible system. Still, the latest result for the
interacting spin system \cite{EB2} gives a square root rather than the
logarithmic behavior.

To the best of our knowledge,
$t_{\hbar}$ has not been yet detected experimentally. On the other hand,
it is known that random external noise \cite{r11,r12} as well as
decoherence \cite{r2,r4,r13,r14} suppress evidence of the breakdown at $t_\hbar$
 or at $T_L$, restoring
quantum-classical correspondence.
This happens when the coherence length \cite{r14,r15}
\begin{equation}
l_c=\hbar \sqrt{\lambda/2D}
\end{equation}
maintained by the chaotic system in the presence of decoherence (which can be
parametrized under certain conditions \cite{r14,r15} by the addition of the
diffusive term $\sim D\partial_p^2W$ to the evolution equation, (\ref{Moyal}))
becomes small compared to the scale of nonlinearities $\chi$
\begin{equation}
l_c\ll \chi
\end{equation}
The price for the restoration of quantum-classical correspondence by
decoherence is irreversibility \cite{r2,r14}.

\begin{figure}[htb]
\centering
{\epsfig{file=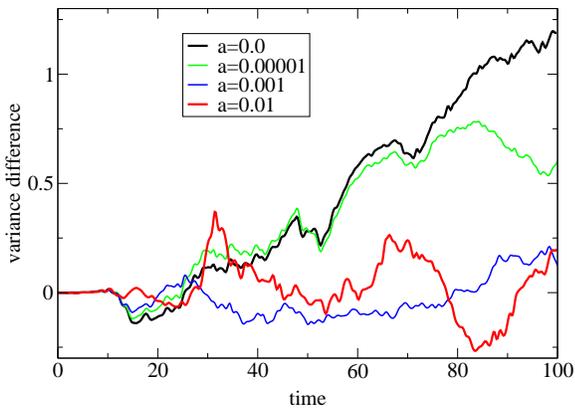, width=8.6cm}}
\caption{Difference between classical and quantum variance for four values of
parameter $a$, as indicated in the figure, and for $\hbar=0.16$ . Addition of harmonic potential
destroys dynamical localization present for $a=0$. All initial conditions are
the same.} 
\label{fig3}
\end{figure}

\begin{figure}[htb]
\centering
{\epsfig{file=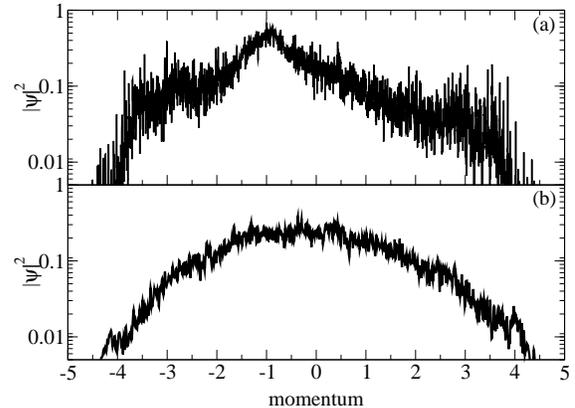, width=8.6cm}}
\caption{Time average probability density in momentum representation:
(a) exponentially localized for $a=0$, (b) Gaussian for $a=0.01$.}
\label{fig4}
\end{figure}

Let us now consider the localization.
At the onset of localization (clearly visible in Fig. \ref{fig3} for $a=0$)
the difference between the classical and quantum variances 
begins to rapidly increase, in a manner approximately consistent with
$\Delta^2_{\mathrm  cl}-\Delta^2_{\mathrm qm} \cong d(t-T_L)$.
 This is to be expected. In systems
effected by localization wave functions exhibit characteristic exponential form,
as exemplified in Fig.~\ref{fig4}(a).

\begin{figure}[htb]
\centering
{\epsfig{file=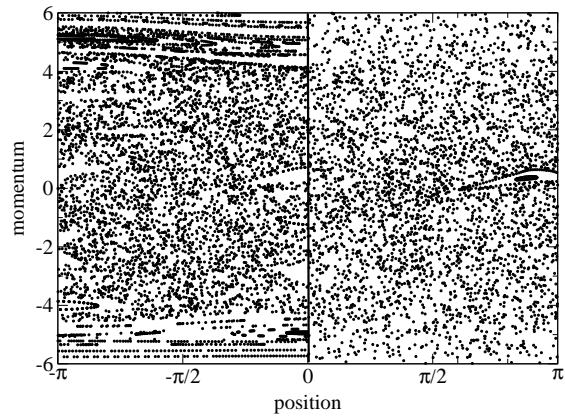, width=8.6cm}}
\caption{Stroboscopic Poincar$\acute {\rm e}$ surface of sections for the system
with parameter $a=0$ (left), and for $a=0.01$ (right).} 
\label{fig5}
\end{figure}

Localization can be suppressed by noise and decoherence, as was recently
shown by two groups using similar experimental setups \cite{r15,r15a,r15b}.
Figure \ref{fig3} shows that a much simpler mechanism for suppression of
localization exists: When the driven periodic potential is supplemented by
even a weak harmonic potential, [controlled by the parameter $a$ in (\ref{e5})]
localization disappears while chaotic character of motion persists,
as illustrated by Poincar$\acute {\rm e}$ surfaces of section in Fig.~\ref{fig5}.

In the presence of even a weak harmonic force 
a prominent growth of the difference
between quantum and classical averages nearly disappears. Also
wave functions cease to be exponentially localized -- compare
Fig.~\ref{fig4}(b). At the first blush, one
may find this dramatic response to even a weak harmonic potential quite
surprising.

Our explanation of this effect is simple and general. Numerical studies show 
that
the suppression of localization occurs when  quarter of the period of the 
harmonic oscillator is approximately the localization time scale $T_L$.
Localization can be maintained only when the "swapping" between $x$ and $p$
generated by the harmonic evolution happens on a time scale longer than $T_L$.
(After all, localization in $p$ is incompatible with localization in $x$,
since the wave function cannot have the exponential form in both complementary
observables.) In fact, we have seen disappearance of localization when the
period of the harmonic oscillator is about $16T_L$. So, the "swapping" does not
need to be very frequent. This simple explanation of suppression of
localization is model independent and may be the key result of our paper.

Experimental verification of our two key results concerning (i) $t_\hbar$ and
(ii) $T_L$ should be possible. It will be more difficult to check for the intriguing prediction of the correspondence
breakdown on short time scale $t_\hbar$. One would ideally want to have 
otherwise
identical two systems, one of them quantum, the other classical. In our
quantum Universe this is unfortunately impossible. One remedy would be
then to compare real experimental results with the classical computer
experiment for the same values of all relevant parameters.

There is, however, something disingenuous about comparing a classical computer
simulation with a quantum experiment in an attempt to find a breakdown of
quantum classical correspondence. We shall therefore suggest an alternative
strategy: Effective value of $\hbar$ can be adjusted in the experiments
\cite{Raizen,r15}. Therefore, it may be possible to look for differences between
some average quantities, say $\Delta_2$, for two otherwise identical systems
which are quantum to a different extent that is adjusted by selecting
different values of $\hbar$. It is obvious from Fig.~\ref{fig1} that significant
differences between two systems endowed with different effective value $\hbar$
would appear, and it should be possible to check for the validity of (\ref{e1}).

Destruction of localization by introduction of the harmonic trap into
experiments which test quantum chaos in optical lattices should
be much easier to verify. It may be desirable to superpose optical lattice of
the sort used to investigate quantum chaos \cite{Raizen,r15} in
a magneto-optical trap anyway (e.g. to confine the system such as BEC,
where investigation of chaos may be of interest in its own right
 \cite{Gardiner}). It
should be then relatively easy to see if our prediction of the loss of
localization is indeed borne out by the experiments.

Both of our key conclusions, while investigated numerically in a specific 
system selected for its relevance to experiment should have a broad validity.
In particular, destruction of localization due to rotation of phase space
as a whole by the background (e.g. harmonic) potential on a time scale
comparable to the localization time $T_L$ is justified by a simple physical
argument. This effect is dramatic and should be easily accessible to
experiments. The earlier and more subtle phenomena that occur at $t_\hbar$
are robust to the addition of an overall potential, providing that the chaotic
character of the system is preserved. Our suggestion for investigation of the
phenomena relies on the ability to adjust the relevant effective $\hbar$ in 
experiments. Moreover, we have found indication that such quantum-classical
discrepancies in the averages diminish more slowly than ${\cal O}(\hbar)$,
which -- if confirmed in experiments we suggest -- would be of major
importance to our understanding of the classical limit of quantum theory.

 JZ and ZPK acknowledge support by KBN under grant 2P03B00915.

\end{multicols}

\end{document}